\documentclass[%
reprint,
 amsmath,amssymb,
 aps,
 prl,
]{revtex4-2}
\usepackage{overpic}
\usepackage{url}
\usepackage{graphicx}% Include figure files
\usepackage{dcolumn}% Align table columns on decimal point
\usepackage{upgreek}
\usepackage{bm}% bold math
\usepackage{color}
\usepackage{hyperref}
\usepackage{multirow} 
\usepackage{booktabs} 
\hypersetup{%hypertex=true,
            colorlinks=true,
            linkcolor=blue,
            anchorcolor=blue,
            citecolor=blue}
\usepackage{siunitx}

\begin{document}

%\preprint{APS/123-QED}

%\title{Energy Stabilization of High-Charge Bunches from Plasma-based Accelerators Using a Beam-driven Active Plasma Dechirper}

\title{Beam-Driven Active Plasma Dechirper for Energy Stabilization of High-Charge Bunches}

\author{Xueyan Shi}
\author{Haisheng Xu}%
 \email{xuhs@ihep.ac.cn}
\author{Dazhang Li}
 \email{lidz@ihep.ac.cn}
\author{Wei Lu}
 \email{weilu@ihep.ac.cn}
\affiliation{Institute of High Energy Physics, Chinese Academy of Sciences, Beijing 100049, China}

%\affiliation{University of Chinese Academy of Sciences, Beijing 101408, China}

\date{\today}% It is always \today, today,

\begin{abstract}
Plasma-based accelerators offer ultrahigh electric field gradients, yet delivering high-charge bunches with simultaneously low energy jitter and narrow energy spread remains a critical challenge. We propose a beam-driven active plasma dechirper integrated with a magnetic chicane to suppress both imperfections. By operating in the blowout regime at low plasma density, the dechirper accommodates a stretched witness bunch, thereby mitigating beam-loading effects and preserving the linear slope of the longitudinal wakefield for precise energy stabilization. Furthermore, a tailored initial energy chirp is introduced to reconcile the disparate optimal plasma lengths required for correcting jitter and spread. Three-dimensional simulations demonstrate that for a 1~GeV, 1~nC electron bunch, this scheme simultaneously reduces the energy jitter and spread from $\pm2$\% and 1\% to $\pm0.05$\% and 0.05\%, respectively, while maintaining nearly 100\% transmission.

\end{abstract}

\maketitle

%\tableofcontents
%\section{Introduction}
Plasma-based accelerators (PBAs)~\cite{TajimaPRL1979,ChenPRL1985} represent a revolutionary advance in particle acceleration, featuring ultrahigh accelerating gradients above $100\ \mathrm{GV\cdot m^{-1}}$—three orders of magnitude beyond conventional radio-frequency (RF) cavities. Global research has reached major milestones over the past decades, producing electron beams with 10~GeV energy~\cite{AniculaeseiMRE2023}, nC-scale charge~\cite{GotzfriedPRX2020,LiPOP2017} and sub-$\rm \upmu mrad$ emittance~\cite{PlateauPRL2012}. Despite these advances, the intrinsic energy jitter and energy spread of PBA beams remain critical obstacles for practical applications such as advanced light sources and next-generation high-energy colliders.

Two complementary routes have been explored to improve the energy stability. The first focuses on {\it in situ} optimization during the injection and acceleration stages, using techniques such as controlled injection~\cite{PakPRL2010,JainSC2024}, optimized beam loading~\cite{TzoufrasPRL2008}, and plasma tailoring~\cite{GonsalvesNPhy2011}. These approaches have yielded permille-level energy spread and percent-level energy jitter for tens-of-pC beams~\cite{WangNATURE2021}, but they generally become less effective for higher charges. The second pursues post-acceleration longitudinal phase space (LPS) manipulation, typically using a magnetic chicane to map energy deviations into timing jitters, after which an active dechirper provides differential energy correction to compensate both jitter and spread~\cite{WinklerNature2025}. This route decouples stability improvement from the acceleration stage and is, in principle, applicable to higher charges.

However, existing active dechirpers face inherent drawbacks. RF-cavity-based dechirpers~\cite{WinklerNature2025} provide a broad linear correction range for large energy deviations but are constrained by low field gradients and the need for dedicated power sources and waveguides. Laser-driven active plasma dechirpers (LAPDs)~\cite{PousaPRL2022,Shi_ipac2022weozgd1} offer higher field strength and have realized 0.1\% level energy stability for tens-of-pC beams. For high-charge beams, however, severe beam loading distorts the driver wakefield, destroying longitudinal field linearity and deteriorating the energy spread. Reducing the plasma density can mitigate beam loading but requires impractical PW-level laser power to satisfy the nonlinear wakefield criterion $a_0 \gtrsim 1$ and the matching condition $k_p w_0 \sim 2\sqrt{a_0}$~\cite{LuPRSTAB2007} (where $a_0$ is the normalized vector potential of the laser, $w_0$ the laser spot size, $k_p$ the plasma wavenumber). As an improved alternative, a two-stage dechirping scheme combining dual magnetic chicanes, an LAPD, and a passive plasma dechirper~\cite{LiuPRL2024} has been proposed~\cite{ShiNJP2024}, achieving 0.1\% energy jitter and 0.5\% energy spread for $\sim$510~pC bunches. Nevertheless, this design suffers from pronounced coherent synchrotron radiation (CSR), chromatic aberrations, and higher-order dispersion in dual-chicane layouts, and the increased complexity compromises both compactness and operational reliability.

\begin{figure*}[!htpb]
    \centering
    \begin{overpic}[width=0.98\textwidth]{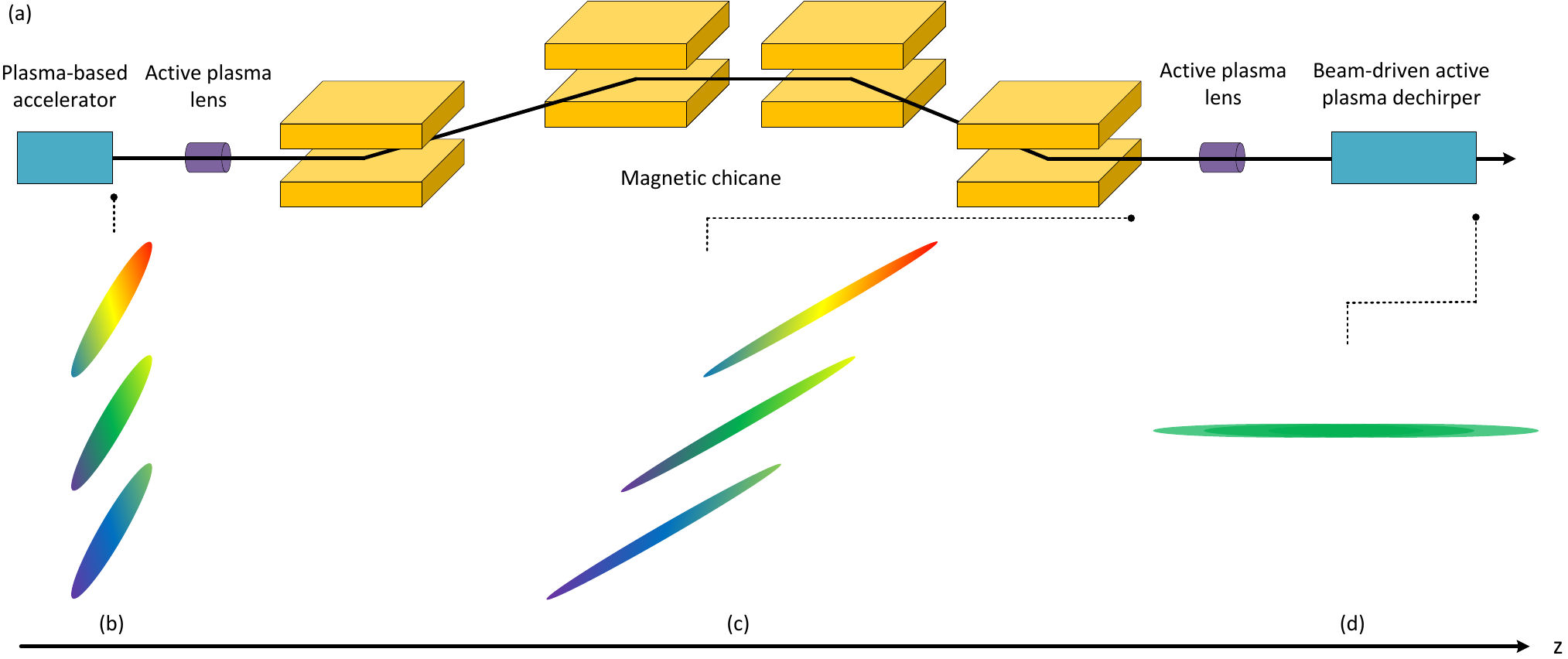}
    % \put(1,60){\textbf{(a)}}
    % \put(8,21){\textbf{(b)}}
    \end{overpic}
    \caption{\label{fig:Fig01} Schematic of the proposed scenario combining a magnetic chicane and a BAPD for simultaneous correction of energy jitter and energy spread in high-charge electron beams. (a) Overall layout of the beamline, consisting of active plasma lenses, a magnetic chicane, and the BAPD. (b)–(d) Evolution of the longitudinal phase space of the bunches with initial energy deviation and energy spread along the beamline, where particles are color-coded by energy (red: high, blue: low).
}
\end{figure*}

In this paper, we propose a beam-driven active plasma dechirper (BAPD) integrated with a magnetic chicane to overcome these limitations for high-charge bunches. With a simple drive beam density condition of $n_b\gtrsim n_p$, the BAPD operates in the blowout regime and enables robust wakefield regulation under high bunch charge. The matching condition $\sigma_r^4=2\epsilon_{nr}^2/(\gamma k_p^2)$~\cite{BarovPRE1994} (where $\sigma_r$ and $\epsilon_{nr}$ are the beam size and normalized emittance of the driver, respectively, and $\gamma$ the Lorentz factor) supports a large transverse beam size at low plasma density, which is readily achievable in experiments. The resultant elongation of the witness bunch length effectively reduces the slice energy spread and mitigates detrimental beam-loading distortions. Furthermore, finite beam loading creates divergent optimal plasma lengths for minimizing energy jitter and spread, a discrepancy that can be fully compensated by introducing a moderate initial linear energy chirp to high-charge witness. Start-to-end simulations verify that using a 0.1~GeV, 4~nC drive beam enables our scheme to reduce the energy jitter and spread of 1~GeV, 1~nC electron beams from $\pm2\%$ and $1\%$ to $\pm0.05\%$ and $0.05\%$, respectively, with nearly 100\% transmission. This provides a feasible and compact path to high-stability, nC-level PBA beams for high-energy accelerator applications.

The proposed scheme is illustrated in Fig.~\ref{fig:Fig01}(a). We consider a high-charge witness electron beam with initial energy $E_0(1\pm\Delta)$, uncorrelated (slice) energy spread $\sigma_{u_0}$, linear energy chirp $h_0$, and rms bunch length $\sigma_{z_0}$, yielding a total energy spread $\sigma_{\delta_0}=h_0 \sigma_{z_0}+\sigma_{u_0}$, where $E_0$ is the reference energy and $\Delta$ is the energy jitter. As the witness beam is emitted directly from the PBA, the shot-to-shot energy jitter is considered as shown in Fig.~\ref{fig:Fig01}(b).

After being focused by an active plasma lens (APL)~\cite{TilborgPRL2015}, the witness enters a four-dipole magnetic chicane, in which the energy deviations are converted into longitudinal position differences, establishing a linear time–energy correlation while stretching the bunch. This energy–position mapping imprints a tunable linear energy chirp and reduces the slice energy spread, as illustrated in Fig.~\ref{fig:Fig01}(c). A second APL downstream refocuses the beam for efficient injection into the plasma dechirper.

The BAPD is driven by a high-charge electron beam and operates in the blowout regime, providing a longitudinal wakefield with a linear slope. By properly positioning the witness within the wakefield, higher-energy particles experience deceleration while lower-energy particles undergo acceleration. This differential energy correction simultaneously suppresses both energy jitter and energy spread, yielding a high-stability, low-energy-spread beam as shown in Fig.~\ref{fig:Fig01}(d).

In practice, the witness itself also excites a wakefield during propagation. Its superposition with the driver-excited wakefield distorts the overall wakefield structure within the witness, known as the beam loading effect, degrading the linearity of dechirping. For low-charge beams, beam loading is negligible, so the optimal plasma interaction lengths for energy jitter correction and energy spread reduction are nearly identical. For high-charge beams, however, the substantial beam loading reduces the effective dechirping gradient and gives rise to distinct optimal lengths for minimizing energy jitter and spread:
\begin{equation}
L_{\rm jitter} = \frac{E_0}{e\mathcal{E}_0^\prime R_{56}},
\label{eq:3}
\end{equation}
\begin{equation}
L_{\rm spread} = \frac{E_0}{e\mathcal{E}_1^\prime h_0}, 
\label{eq:4}
\end{equation}
where $\mathcal{E}_1^\prime$ is the slope of the combined wakefield. Given that $\mathcal{E}_0^\prime>\mathcal{E}_1^\prime$, an initial linear energy chirp is applied to the witness:
\begin{equation}
h_0 = \left[R_{56}\left(\frac{\mathcal{E}_0^\prime}{\mathcal{E}_1^\prime}-1\right)\right]^{-1}. 
\label{eq:5}
\end{equation}
This chirp compensates the beam loading effect and equalizes the two optimal interaction lengths.

The theoretical minimum uncorrelated energy spread $\sigma_{u_1}$ is inversely proportional to $R_{56}$. Assuming the witness occupies a phase region $\lesssim\lambda_p/2$ in the BAPD, with $\lambda_p=2\pi/k_p$ the plasma wavelength, the optimum $R_{56}$ can be estimated as
\begin{equation}
R_{56} \approx \frac{\lambda_p}{4(\Delta+4\sigma_{\delta_0})}. 
\label{eq:6}
\end{equation}
For a symmetric four-dipole chicane, $R_{56}=4L_m\left[\sin\theta/\theta-\cos\theta\right]+2L_d\theta\sin\theta$~\cite{WChao2013}; $L_d$ is the drift length between dipoles, $\theta$ and $L_m$ are the bending angle and length of each dipole, respectively.

Start-to-end simulations are performed to validate the proposed scheme. Conventional beamline elements (drifts, magnets) are modeled with the particle-tracking code {\tt ELEGANT}~\cite{osti_761286}, including the 1D CSR effect. The BAPD is simulated using the fully 3D quasi-static particle-in-cell (PIC) code QuickPIC~\cite{HuangJPCS2006}.  APLs are modeled with the quasi-3D PIC code FBPIC~\cite{LEHECPC2016}.

\begin{figure}[!htpb]
    \centering
    \begin{minipage}{0.99\linewidth}
    \centering
    \includegraphics[width=0.99\columnwidth]{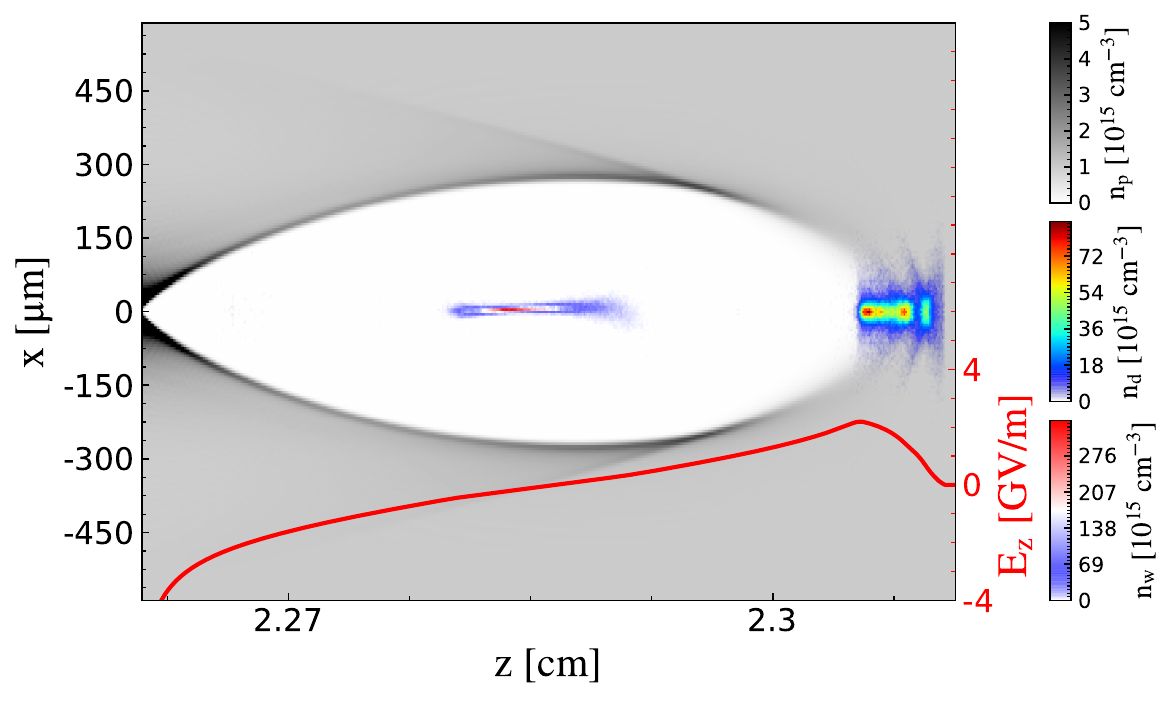}
    \end{minipage}
%    \put(-450,120){\textbf{(a)}}
    \put(-240,60){\textbf{(a)}}
    \qquad
    \begin{minipage}{0.98\linewidth}
    \centering
    \includegraphics[width=0.98\columnwidth]{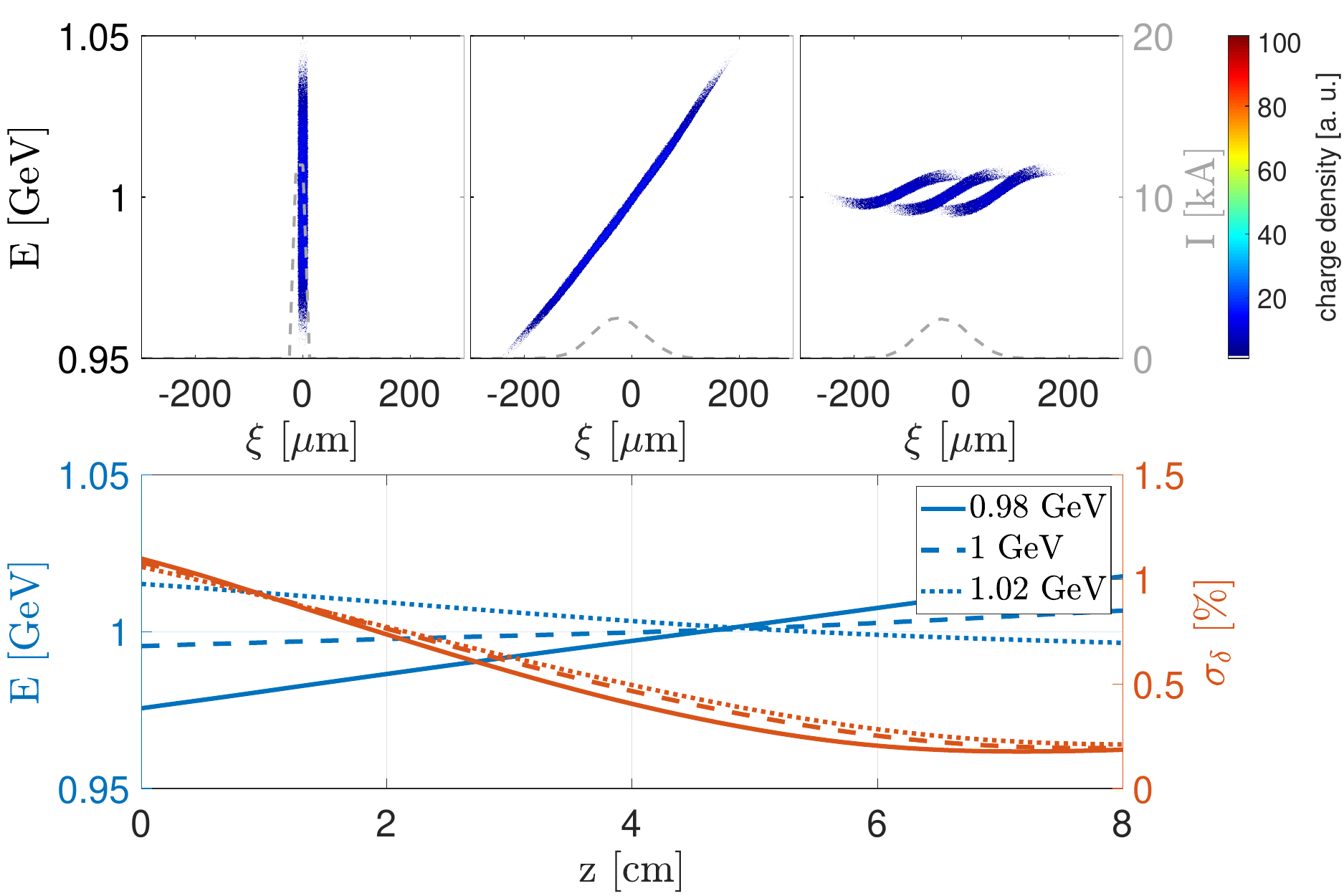}
    \end{minipage}
    %\put(-405,127){\textbf{(b)}}
    %\put(-295,127){\textbf{(c)}}
    %\put(-185,127){\textbf{(d)}}
    %\put(-405,-17){\textbf{(e)}}
     \put(-212,65){\textbf{(b)}}
     \put(-154,65){\textbf{(c)}}
     \put(-96,65){\textbf{(d)}}
     \put(-212,-12){\textbf{(e)}}
    \qquad   
    \caption{\label{fig:Fig02_1}(a) Beam and plasma profiles at the BAPD center. (b)–(d) Longitudinal phase space evolution at the PBA exit, chicane exit, and BAPD exit, with beam current overlaid (gray). (e) Central energy and energy spread evolution along the BAPD.}
\end{figure}

A flat-top electron beam with Gaussian transverse profiles serves as the driver, while a flat-top beam with Gaussian head/tail ramps is used as the witness. Key parameters include: witness energy $E_0=1\ \text{GeV}$, energy jitter $\Delta=\pm2\%$, initial energy spread $\sigma_{\delta_0}=\sigma_{u_0}=1\%$, charge $Q_{\rm w}=1\ \text{nC}$, rms bunch length $\sigma_{z_0}=5.1\ \mu\text{m}$, and normalized emittance $4\ \mu\text{mrad}$. The driver has an energy of 0.1~GeV, charge $Q_{\rm d}=4\ \text{nC}$, normalized emittance $20\ \mu\text{mrad}$, rms bunch length $102.5\ \mu\text{m}$, and peak density $n_b=9.75\times10^{15}\ \text{cm}^{-3}$. The chicane has $R_{56}=4.4\ \text{mm}$ and a total length 2~m. The BAPD uses a uniform plasma density of $n_p=1\times10^{15}\ \text{cm}^{-3}$, resulting in a normalized current $\Lambda=1.4$ and a normalized charge $\tilde{Q}=13.2$~\cite{LindstrømARXIV2025}, which confirms operation in the blowout regime. The APLs have density $1\times10^{14}\ \text{cm}^{-3}$, length 1~cm, and focusing gradients up to $2330\ \text{T/m}$.

\begin{table*}[!htp]
    \centering
    \caption{\label{tab:table1} Initial and final parameters of the driver and witness beams.}
    \begin{ruledtabular}
    \begin{tabular}{lccccc}
    \multirow{2}{*}{\textbf{Parameter}} & \multirow{2}{*}{\textbf{Unit}} & \multicolumn{2}{c}{\textbf{Driver}} &
    \multicolumn{2}{c}{\textbf{Witness}} \\
    \cmidrule{3-4}
    \cmidrule{5-6}
    ~ & ~ & \textbf{BAPD entrance} & \textbf{BAPD exit} & \textbf{BAPD entrance} & \textbf{BAPD exit} \\
    \colrule
    Energy $E$      & MeV      & 100.0 & 41.3 & 999.7 & 1000.2 \\
    Charge $Q$      & nC       & 4.0   & 4.0  & 1.0   & 1.0 \\
    Bunch length $\sigma_z$ & $\upmu$m & 104.3 & 127.3 & 62.4  & 62.4 \\
%    Beam size $\sigma_{r}$ & $\upmu$m & 31.5 &  & 4.0 & 5.8/5.8 & 6.0/21.6 \\
    Norm. emittance x/y & $\upmu \rm m\ rad$ & 20.0/20.0 & 66.2/66.9 & 21.4/11.1 & 22.2/11.0 \\
    Energy spread $\sigma_\delta$ & \% & 20.0 & 33.7 & 1.10 & 0.04 \\
    Slice energy spread $\sigma_u$ & \% & 1.0 & 6.5 & 0.03 & 0.03
    \end{tabular}
    \end{ruledtabular}
\end{table*}
%We first perform the case where the witness without an initial energy chirp. The beams are focused by the APL and stretched by the chicane, where the normalized emittance increases from $5.7\ \rm \upmu mrad$ to $6.7\ \rm \upmu mrad$ due to the CSR effect (which could be suppressed with alternative chicane designs). Then they enter the BAPD for energy compression. When the charge density of the BAPD driver is sufficiently higher than the plasma density (here $n_d\gg n_p$), the background electrons will be fully expelled by the space charge force and a trailing wakefield with nearly uniform $\mathcal{E}_0^\prime$ is generated, as shown in FIG.~\ref{fig:Fig02_1}(a). FIG.~\ref{fig:Fig02_1}(b) illustrates the LPS evolution of the witness beams at the beginning, the chicane exit, and the BAPD exit, respectively. It can be seen that the wakefield generated by the driver effectively corrected the energy deviations and reduced the energy chirp simultaneously. The evolution of the energy and energy spread of the witness beams throughout the BAPD is plotted in FIG.~\ref{fig:Fig02_1}(e). Clearly, the energy spreads reach the minimum values of $\sigma_{u_1}\sim0.12\%$ after $L_{\rm spread}=6.7\ \rm cm$, while the energy jitters reach the minimum values of 0.05\% with a residual energy chirp after $L_{\rm jitter}=4.9\ \rm cm$, see FIG.~\ref{fig:Fig02_1}(d). This length mismatch arises from the nonlinearities in $E_z$ induced by the beam loading, as well as driver diffusion and energy loss during propagation, which collectively mitigate energy chirp removal.

Figure~\ref{fig:Fig02_1}(a) shows a snapshot of the driver, witness, and longitudinal wakefield at the BAPD center. The driver fully expels plasma electrons to form a wakefield with a near linear slope for dechirping. Figures~\ref{fig:Fig02_1}(b)–(d) display the LPS evolution at the PBA exit, chicane exit, and BAPD exit, respectively. The chicane establishes a time–energy correlation, and the BAPD differentially corrects energy deviations across the bunches. Figure~\ref{fig:Fig02_1}(e) tracks the central energy and energy spread along the BAPD. Without an initial chirp, the energy spread reaches the minimum of $\sigma_{u_1}\sim0.12\%$ at $L_{\rm spread}=6.7\ \text{cm}$, whereas the energy jitter is minimized to $0.05\%$ at $L_{\rm jitter}=4.9\ \text{cm}$, a mismatch that is induced by the beam loading. 
%This mismatch arises from beam-loading-induced wakefield distortion and driver evolution, which together prevent simultaneous minimization of the energy jitter and energy spread at a single plasma length.

\begin{figure}[!b]
    \centering
    \begin{minipage}{0.98\linewidth}
    \centering
    \includegraphics[width=0.98\columnwidth]{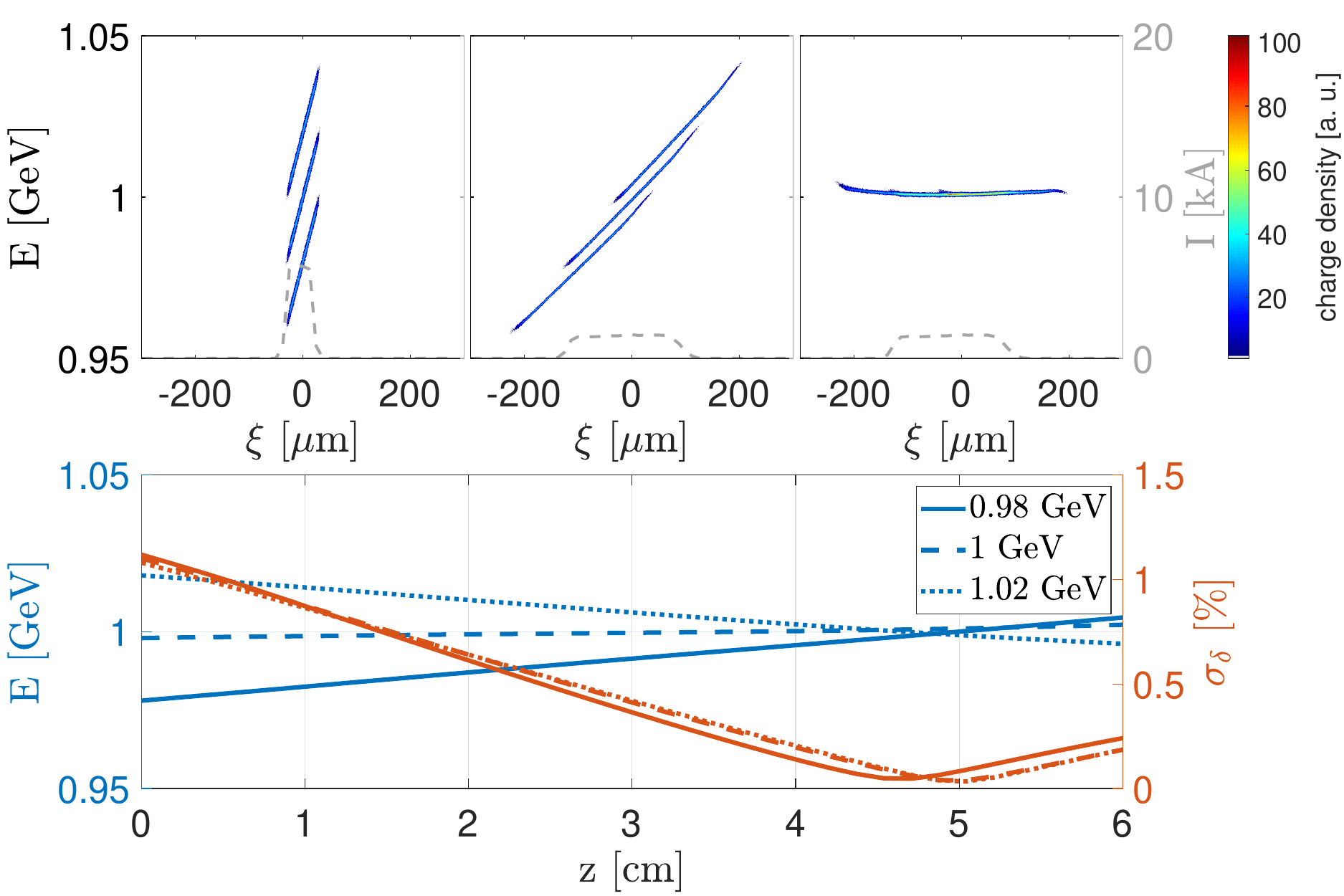}
    \end{minipage}
    %\put(-405,127){\textbf{(a)}}
    %\put(-295,127){\textbf{(b)}}
    %\put(-185,127){\textbf{(c)}}
    %\put(-405,-17){\textbf{(d)}}
     \put(-213,66){\textbf{(a)}}
     \put(-155,66){\textbf{(b)}}
     \put(-97,66){\textbf{(c)}}
     \put(-213,-11){\textbf{(d)}}
    % \qquad
    % \begin{minipage}{0.98\linewidth}
    % \centering
    % \includegraphics[width=0.98\columnwidth]{figure02_5.pdf}
    % \end{minipage}
    % \put(-230,60){\textbf{(d)}} 
    \caption{\label{fig:Fig02_2} Evolution of the high-charge beams in the longitudinal phase space $\left(\xi,E\right)$ for 3 cases with different central energies and a initial energy chirp of $h_0=650\ \rm m^{-1}$, at (a) the PBA exit, (b) the chicane exit, (c) the BAPD exit. The gray dashed lines indicate the current profile of the reference beam. (d) Evolution of the energy and energy spread in the BAPD.}
\end{figure}

To remove the residual chirp, the witness is imparted with an initial chirp, as illustrated in FIG.~\ref{fig:Fig02_2}(a)-(b). To maintain a constant initial total energy spread, the initial bunch length and slice energy spread are correspondingly adjusted to $\sigma_{z_0}=15\ \rm \upmu m$ and 0.1\%. The focusing gradient of the first APL is tuned to $2180\ \rm T\ m^{-1}$. In the case without an initial energy chirp, Eq.~(\ref{eq:3}) and Eq.~(\ref{eq:4}) yield $\mathcal{E}_0^{\prime}\simeq 4.7\ \rm GV\ mm^{-1}\ m^{-1}$ and $\mathcal{E}_1^{\prime}\approx 3.4\ \rm GV\ mm^{-1}\ m^{-1}$, which implies an initial chirp of $h_0=600\ \rm m^{-1}$ according to Eq.~(\ref{eq:5}). Accounting for variations in the initial witness parameters alongside the head-to-tail nonuniform and nonlinear evolution of $\mathcal{E}_1^{\prime}$, this chirp is further optimized to $650\ \rm m^{-1}$ in simulations.

Figs.~\ref{fig:Fig02_2}(a)–(c) display the LPS evolution of the witness beams with an initial chirp, and Fig.~\ref{fig:Fig02_2}(d) shows the central energy and energy spread in the BAPD. Using the same BAPD parameters as before, both the energy jitter and spread drop to a minimum of 0.05\% after a plasma length of 4.9~cm, with detailed parameters listed in TABLE~\ref{tab:table1}. The peak current of the witness at the BAPD exit is $\sim$1.5~kA for a bunch length of $62.4\ \rm \upmu m$. The final normalized emittances of $22.4\ \rm \upmu mrad$ increase mostly due to chromatic effects and CSR, which can be further suppressed by optimizing the APL profiles or employing sextupole magnets within the chicane. Start-to-end simulations for Gaussian and triangular electron beams show similar results, demonstrating the robustness and feasibility of the scheme.

The preceding analysis assumes a linear initial chirp of the witness, which is generally valid for PBA beams. For a nonlinear initial chirp, two issues arise. First, the linear wakefield generated by the BAPD compensates only the linear chirp, leaving nonlinear correlations unaffected. However, these higher-order correlations are expected to be weak because the low plasma density requires a large $R_{56}$, making the post-chicane chirp predominantly linear. Second, and more importantly, a nonlinear initial chirp prevents precise compensation of the mismatch between $L_{\rm jitter}$ and $L_{\rm spread}$. Suppressing higher-order chirp components requires nonlinear LPS manipulation, which remains to be investigated.

%\section{Brief Comparison between BAPD and LAPD}

We also evaluate the performance of the BAPD relative to the LAPD. Since both schemes can effectively reduce energy jitter regardless of the witness charge, the comparison is focused on energy spread reduction. For a consistent benchmark, the witness parameters are set to $E_0=500\ \rm MeV$, $\sigma_{\delta_0}=1.2\%$, and $Q_{\rm w}=510\ \rm pC$, the same as the values used in Ref.~\cite{ShiNJP2024}. The dechirpers are driven under varying plasma densities using particle drivers (with charges of $Q_{\rm d}=2\ \rm nC,\ 4\ nC,\ 6\ nC$) and laser drivers (with peak power of $P=200\ \rm TW,\ 500\ TW,\ 1000\ TW$). All the particle drivers have a 100~MeV energy, a 20\% energy spread, and a $100\ \rm \upmu m$ bunch length, while all the laser drivers have a 25~fs FWHM duration, a $w_0=25\ \rm \upmu m$, and corresponding $a_0=3.1,\ 4.9,\ 6.9$. PIC simulations for the LAPD and BAPD are conducted using the codes FBPIC and QuickPIC, respectively.

\begin{figure}[!b]
    \centering
    \begin{overpic}[width=0.49\textwidth]{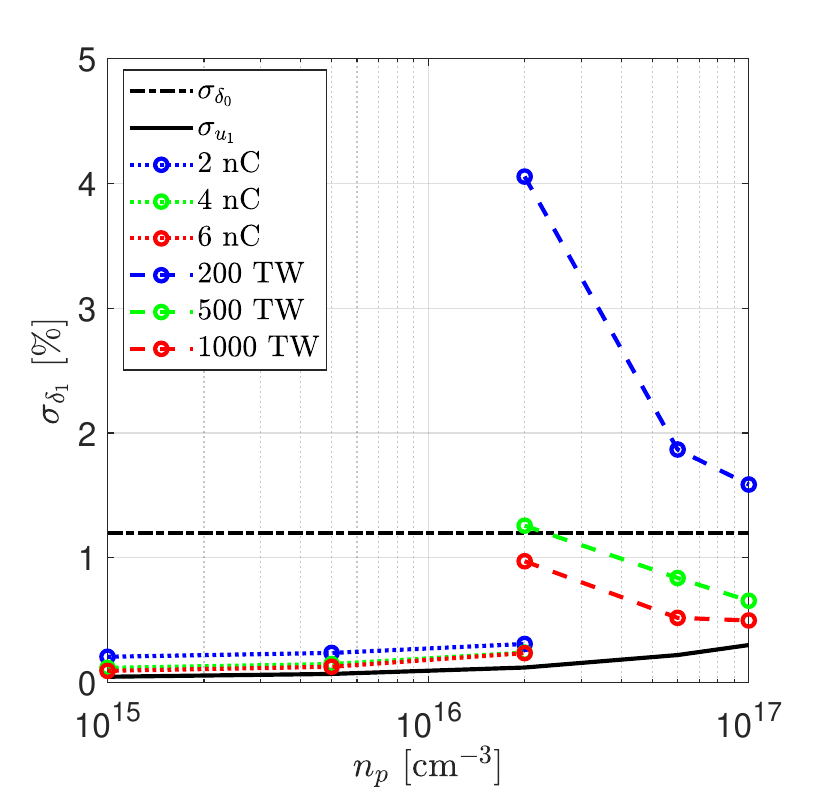}
    \end{overpic}
    \caption{\label{fig:Fig03} Comparison of the energy spread obtained by the BAPD and LAPD for an electron beam with a 500~MeV energy and 510~pC charge. In the legend, the charge and power denote the bunch driver charge and laser peak power, respectively.
}
\end{figure}

Figure~\ref{fig:Fig03} shows the final energy spreads $\sigma_{\delta_1}$ for both schemes across different plasma densities. It can be seen that the LAPD scheme requires a 1~PW laser to reduce the energy spread of a 510~pC beam from 1.2\% to 0.5\%, a performance comparable to that reported in Ref.~\cite{ShiNJP2024}, which also highlights the challenge of scaling to higher charges. Although lowering $n_p$ decreases $\sigma_{u_1}$ and mitigates beam loading, ensuring stable laser propagation necessitates high laser power, and simulations reveal that the concurrent reduction in $\mathcal{E}_1^\prime$ is more significant, ultimately leading to an increased $\sigma_{\delta_1}$ (dashed lines). Conversely, the BAPD scheme reduces the energy spread to below 0.3\% utilizing only a 100~MeV electron driver (dotted lines). As the driver charge increases and the plasma density decreases, a lower energy spread of $\sigma_{\delta_1}<0.1\%$ is achievable. It is worth noting that the BAPD requires only a high-charge, high-current driver, without imposing stringent constraints on its energy, energy spread, energy chirp, or temporal profile. Such bunches can be readily generated either by 100~TW-class lasers via laser wakefield acceleration~\cite{LuPRSTAB2007} or by compact RF accelerators. In other words, under equivalent laser power, the BAPD exhibits superior performance in energy stabilization compared to the LAPD.

%\section{Conclusion}

In summary, we have implemented simulations to show a beam-driven active plasma dechirper combined with a magnetic chicane that, for the first time, simultaneously suppresses the energy jitter and total energy spread of 1~nC electron bunches to 0.05\%. The BAPD driven by a high-charge electron bunch can work in the blowout regime at low plasma density, effectively mitigating the beam-loading-induced field distortion, hence inhibiting the energy spread growth. A precisely tuned initial energy chirp is employed to compensate the unmatched optimal plasma lengths for energy jitter and spread correction due to residual beam loading. The achieved dechirping field gradient of $\sim 4.7\ \mathrm{GV\,mm^{-1}\,m^{-1}}$ outperforms conventional RF technology by two orders of magnitude, and the energy spread reduction represents more than an order-of-magnitude improvement over state-of-the-art laser-driven dechirpers for high-charge bunches. Our scheme provides a pathway to meeting the stringent beam quality requirements of future compact colliders and advanced light sources.

%In summary, we have demonstrated a robust scheme for the energy stabilization of nC-level electron beams utilizing a beam-driven active plasma dechirper integrated with a magnetic chicane. By exploiting the wakefield tailored in the blowout regime, this approach effectively overcomes the beam-loading limitations inherent in conventional dechirping methods for high-charge bunches. Simulations indicate that for a 1~GeV, 1~nC beam, the energy jitter and spread are simultaneously suppressed from $\pm2.0\%$ and 1.0\% to approximately $\pm0.05\%$ and 0.05\%, respectively, maintaining near 100\% transmission efficiency. notably, the achieved dechirping strength of $\sim 4.7~\text{GV mm}^{-1}\text{m}^{-1}$ exceeds that of conventional rf technology by two orders of magnitude. Furthermore, compared to laser-driven plasma dechirpers, our scheme provides an order-of-magnitude enhancement in energy spread suppression for high-charge regimes. With further optimization of the driver and witness parameters, performance is expected to improve further. These results mark a critical step toward meeting the stringent beam quality requirements of next-generation compact high-energy physics facilities and advanced light sources.

\begin{acknowledgments}
This work is supported by the Strategic Priority Research Program of the Chinese Academy of Sciences, Grant No.~XDB0530000, the National Natural Science Foundation of China (No.~12375149), the National Key R\&D Program of China (Grant No.~2022YFA1603401), and the Research Foundation of Institute of High Energy Physics, Chinese Academy of Sciences (Grant No.~E65452U810).
\end{acknowledgments}

\nocite{*}

\bibliography{main}% Produces the bibliography via BibTeX.

\end{document}